\documentclass[pra,twocolumn,amsmath,amssymb]{revtex4}
\usepackage{graphicx}

\begin{document}

\title{Experimental Quantum Key Distribution at 1.3 Gbit/s Secret-Key Rate over a 10-dB-Loss Channel}

\author{Zheshen Zhang$^{1,2}$}
\author{Changchen Chen$^1$}
\author{Quntao Zhuang$^{1,3}$ }
\author{Franco N. C. Wong$^1$}
\author{Jeffrey H. Shapiro$^1$}

\affiliation{$^1$Research Laboratory of Electronics, Massachusetts Institute of Technology, Cambridge, MA 02139, USA}
\affiliation{$^2$Department of Materials Science and Engineering, University of Arizona, Tucson, AZ 85721, USA}
\affiliation{$^3$Department of Physics, Massachusetts Institute of Technology, Cambridge, MA 02139, USA}

\date{\today}

\begin{abstract}
Quantum key distribution (QKD) enables unconditionally secure communication ensured by the laws of physics, opening a promising route to security infrastructure for the coming age of quantum computers. QKD's demonstrated secret-key rates (SKRs), however, fall far short of the gigabit-per-second rates of classical communication, hindering QKD's widespread deployment. QKD's low SKRs are largely due to existing single-photon-based protocols' vulnerability to channel loss. Floodlight QKD (FL-QKD) boosts SKR by transmitting many photons per encoding, while offering security against collective attacks. Here, we report an FL-QKD experiment operating at a 1.3 Gbit/s SKR over a 10-dB-loss channel. To the best of our knowledge, this is the first QKD demonstration that achieves a gigabit-per-second-class SKR, representing a critical advance toward high-rate QKD at metropolitan-area distances.
\end{abstract}

\maketitle

Shor's algorithm \cite{Shor1997} for polynomial-time prime factorization on a quantum computer poses an existential threat to public-key cryptography, and hence to present-day Internet commerce.  Quantum key distribution (QKD) \cite{Bennett1984} provides means to ward off this threat by enabling remote parties (Alice and Bob) to establish a secret key, comprised of a shared set of random bits, with security that is guaranteed by the laws of physics.  Using that key as a one-time pad (OTP), Alice and Bob can then exchange messages with information-theoretic security.  The appeal of unconditionally secure communication spurred the recent demonstration of ground-satellite QKD \cite{Liao2017}, a first step toward a global quantum-communication network. A major barrier to QKD's widespread deployment, however, is QKD systems' low secret-key rates (SKRs).  For the predominant decoy-state BB84 protocol, which transmits single-photon-level signals to ensure security, channel loss leads to state-of-the-art SKRs in the megabit-per-second class over metropolitan-distance fiber connections \cite{Comandar2014}.  Channel loss has a similar effect on traditional continuous-variable QKD (CV-QKD), whose state-of-the-art SKRs \cite{Huang2015} are lower than those of BB84.  These systems are thus incapable of supporting OTP encryption for gigabit-per-second Internet traffic without dramatic increases in their SKRs, such as might be obtained, e.g., by developing fast and high-efficiency single-photon detectors \cite{Marsili2013}, introducing wavelength-division multiplexing \cite{Yoshino2013}, and employing ultra-low-loss fibers \cite{Korzh2015}.  Even so, their use of single-mode encodings imposes the fundamental $-\log_2(1-\eta)$ bits-per-optical-mode SKR bound for a transmissivity-$\eta$ channel \cite{Takeoka2014,Pirandola2017} on their SKRs in bits/sec. 

Floodlight QKD (FL-QKD) is a radically different protocol that was shown, theoretically, to offer gigabit-per-second-class SKRs over metropolitan-area distances using only available technology \cite{Zhuang2016}.  It does so by transmitting many photons per bit---just as is done in classical communication---while obeying the rate-loss bound on SKR in \emph{bits/mode} by using multi-mode encoding to achieve its high SKR in \emph{bits/sec}.  We previously reported an FL-QKD experiment operating at 55 Mbit/s SKR over a 10-dB-loss channel with security against collective attacks \cite{Zhang2017}, representing a $\sim$50-fold improvement over state-of-the-art. Here, we report an FL-QKD experiment that further boosts the SKR by $\sim$20 times, achieving a 1.3\,Gbit/s SKR over a 10-dB-loss channel. Our experiment, to the best of our knowledge, is the first QKD system operating at a gigabit-per-second SKR.

\begin{figure*}[tbh]
\centering
\includegraphics[width=5.95in]{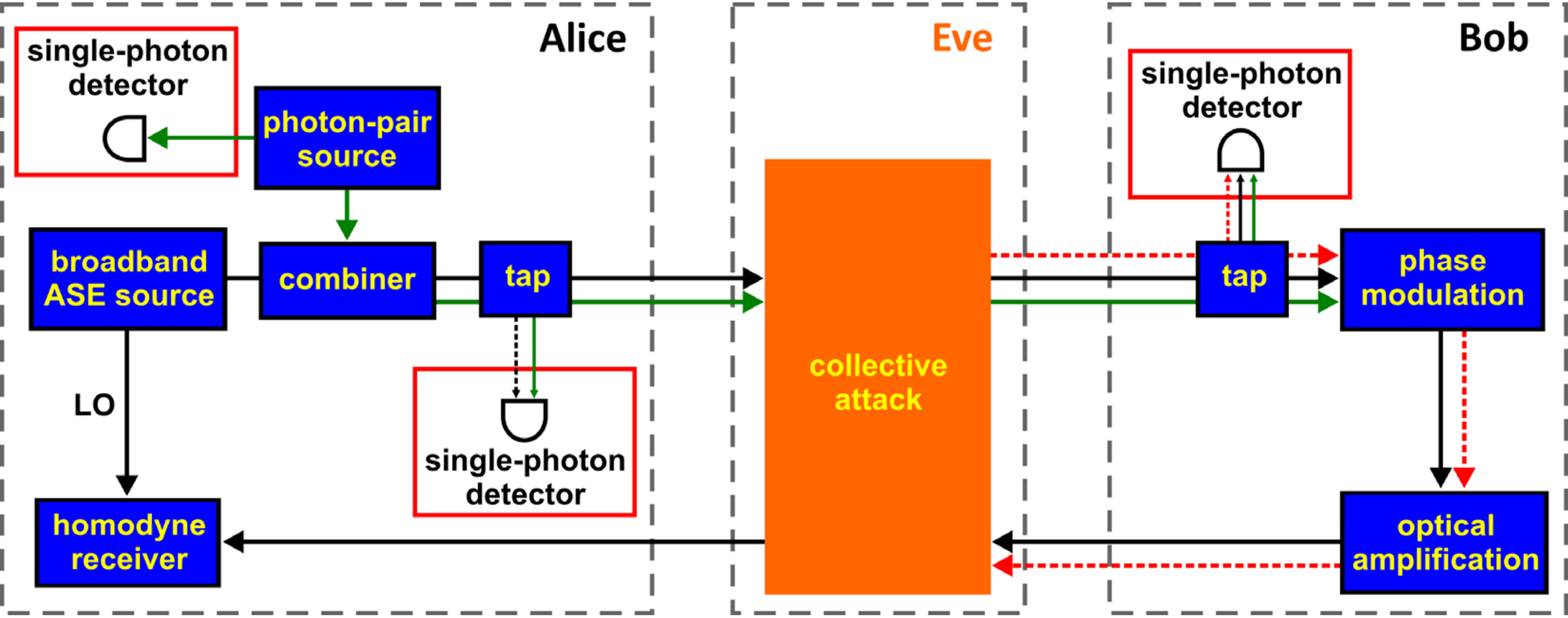}
\caption{\label{fig:FL-QKD_protocol} Schematic of FL-QKD under Eve's optimum collective attack. Light from Alice's broadband (SPDC) source is marked by solid black (green) lines; Eve's injected light is marked by dashed red lines. The channel monitor consists of the three single-photon detectors in red boxes. Only Alice's SPDC photons (solid green) at Bob's detector are coincident with her idler photons, whereas  Eve's injected photons (dashed red) contribute only to the noise background in coincidence measurements. As such, Eve's injection fraction $f_E$ at Bob's terminal can be derived by measuring the coincidences versus singles rates at Alice and Bob's detectors.  Not shown:  classical channel used for reconciliation.}
\end{figure*}
FL-QKD, illustrated in Fig.~\ref{fig:FL-QKD_protocol}, is a two-way continuous-variable protocol that leverages a broadband incoherent source for key distribution, whose optical bandwidth $W$ greatly exceeds the inverse of its encoding time $T$, so that each bit is carried on multiple photons that are spread over $M = TW \gg 1$ optical modes.  It also employs a photon-pair source, together with three single-photon detectors, that enable Alice and Bob to quantify quantum-channel intrusion by their adversary (Eve).  In particular, Alice utilizes an amplified spontaneous emission (ASE) source to generate high-brightness ($\gg 1 $\,photon/mode) broadband light. She then splits the broadband light using a highly-reflecting beam splitter and retains the light from the reflecting arm for use as her homodyne receiver's reference beam. In addition to the broadband classical source, Alice uses a spontaneous parametric down-converter (SPDC) to generate time-correlated signal and idler photons. She immediately detects and time tags her SPDC's idler photons using a single-photon detector. She makes her SPDC's signal photons, which obey thermal statistics, indistinguishable from her ASE photons by forcing their spectra and polarizations to be identical.  Alice combines her SPDC's signal photons with the ASE photons from the weakly-transmitting arm of her initial beam splitter on another unbalanced beam splitter.   Next, she taps, detects, and time tags a small portion of her mixed beam using a second single-photon detector. Alice's two single-photon detectors comprise her part of the channel-monitor system that she and Bob use to quantify Eve's intrusion on their quantum channel, as explained below. Alice transmits the rest of her mixed beam---which consists of low-brightness ($\ll 1$\,photon/mode) ASE light containing much more than 1 photon per bit duration $T$, plus much lower brightness SPDC signal light---to Bob through a quantum channel that is completely under Eve's control. Note that no information has been encoded at this stage. 

Bob first detects and time tags a small portion of the light he receives using a beam splitter and his own single-photon detector, which constitute his part of the channel monitor.  Bob then applies binary phase-shift keying---0\,Rad or $\pi$\,Rad phase shift at rate $R = 1/T \ll W$ bits/sec---to impose random bit-stream modulation on the rest of the light. Next, he amplifies his encoded light with a high-gain optical amplifier, which brings with it three major benefits \cite{Zhuang2016,Zhang2017}: (1) it compensates for the return-path loss to Alice, so despite FL-QKD's being a two-way protocol its SKR only suffers the propagation loss of the Alice-to-Bob path; (2) it adds high-brightness noise to Bob's amplified encoded light, completely masking his random bit-string from a passive eavesdropper; and (3) it obviates Alice's need to have a shot-noise limited homodyne detector---something that traditional CV-QKD requires---because the added noise greatly exceeds the shot-noise level.  Bob sends his encoded and amplified light to Alice, through a channel controlled by Eve, and Alice decodes his random bit-stream with a homodyne receiver that uses her retained reference beam as its local oscillator (LO). 

Eve is assumed to collect all the light that was lost in the Alice-to-Bob and Bob-to-Alice channels. Eve's passive-eavesdropping attack, in which she just makes use of that lost light, is rendered impotent by the no-cloning theorem, which precludes her from regenerating Alice's high-brightness reference from the low-brightness version she captured from the Alice-to-Bob channel.  Without such a reference, Eve cannot penetrate the amplifier noise in which Bob's bit-stream is buried.  A two-way QKD protocol, like FL-QKD, is vulnerable to an active attack \cite{Bostrom2002,Deng2004,Pirandola2008,Shapiro2009,Zhang2013,Shapiro2014,Weedbrook2014}, in which Eve injects some of her own light into Bob's terminal, and decodes Bob's bit-stream using her own stored reference. FL-QKD's SPDC and its three single-photon detectors, however, provide sufficient channel monitoring to quantify and hence defeat Eve's active attack. As illustrated in Fig. \ref{fig:FL-QKD_protocol}, the coincidences rates between Alice and Bob's tapped photons and the idler photons are measured using the three single-photon detectors embedded in red boxes. Only the green signal beams contain photons from the SPDC source that lead to coincidences events with the idler photons. Eve's photons marked in red are uncorrelated with the SPDC's photons and thus only contribute to the noise background in the coincidence measurements \cite{Zhang2017}. As such, the singles versus coincidences rates observed by the three single-photon detectors not only unveil Eve's intrusion, but also quantify the fraction $f_E$ of light reaching Bob's terminal that she injected. Knowledge of $f_E$ allows Alice and Bob to put an upper bound on Eve's Holevo information. Ref.~\cite{Zhuang2016} shows that the active injection attack is the most powerful collective Gaussian attack and, more generally, represents the most powerful collective attack in the frequency domain. A more recent theoretical study \cite{Zhuang2017} elaborates the use of limited entanglement-assisted channel capacity to prove that Gaussian attacks are the optimum for a broad class of two-way QKD protocols. 

\begin{figure*}
\centering
\includegraphics[width=6in]{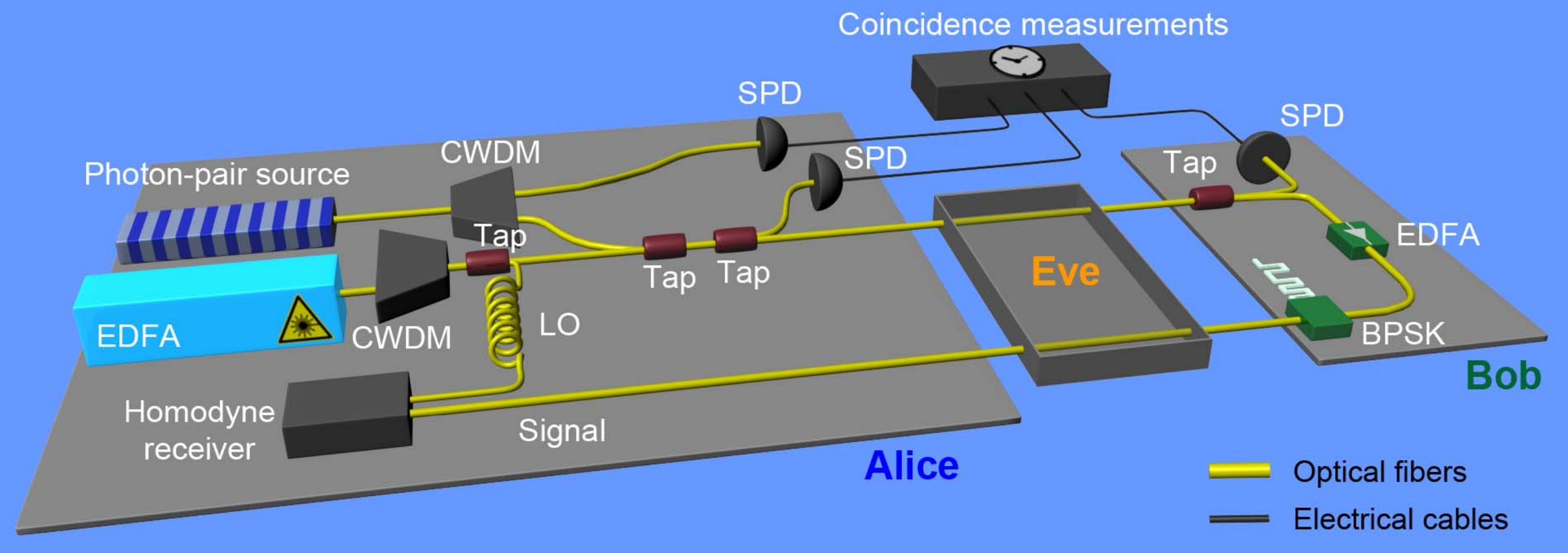}
\caption{\label{fig:exp} Experimental setup. EDFA: erbium-doped fiber amplifier; CWDM: coarse wavelength-division multiplexer; BPSK: binary-phase shift keying; LO: local oscillator; SPD: single-photon detector.  There was no light injection by Eve.}
\end{figure*}

Figure~\ref{fig:exp} shows a schematic of our experiment. An unseeded erbium-doped fiber amplifier (EDFA) emits $\sim$200\,mW of unpolarized broadband ASE light. We first polarize the ASE light using an inline polarizer and filter it with a coarse wavelength-division multiplexer (CWDM) into a $\sim$18-nm (2.24\,THz) band centered at 1550\,nm, before splitting it into two portions with a 90:10 coupler.  The strong ASE portion is retained by Alice in a spool of fibers (including some dispersion-compensating fiber) to serve as the LO for her homodyne receiver. The weak ASE portion from the coupler output is further attenuated by a variable optical attenuator (not shown) and then spectrally shaped using a programmable optical filter (Finisar Waveshaper 1000S, not shown) that modifies its spectrum so that it becomes indistinguishable from that of the SPDC's light. Moreover, the WaveShaper imposes frequency-dependent phase shifts to the ASE light to fine tune its dispersion characteristics. This technique is particularly useful for compensating high-order dispersion that cannot be effectively corrected by conventional dispersion-compensating fibers. The ASE light is later combined with the SPDC's signal light for transmission to Bob.

The SPDC source is a type-0 phase-matched MgO-doped periodically-poled lithium niobate (MgO:PPLN) crystal pumped by a $\sim$1-mW 780-nm beam, producing signal and idler photons centered at 1550\,nm and 1570\,nm, respectively. After the crystal the pump is blocked by two long-pass dichroic filters (not shown), and the signal and idler photons are coupled into a single-mode fiber and then separated with a CWDM filter, resulting in 1.28\,pW signal and idler beams.  The idler photons are immediately detected by a superconducting nanowire single-photon detector (SNSPD) and their detection times are recorded.  A 98:2 fiber tap combines the signal photons (via the 98\% port) and the weak ASE light (via the 2\% port).  Alice taps 2\% of the combined light, detects the tapped photons with a second SNSPD, and records their detection times. She sends the remaining 98\% of the combined light to Bob through a fiber channel with 10-dB attenuation. 

Bob first taps 1\% from the received broadband light, detects photons in that tapped light using a third SNSPD, and records the detection times. The remaining light passes through a circulator (not shown) to block back-propagating light, a CWDM filter (not shown) to reject out-of-band signal, and is then polarization tuned via a paddle (not shown).  An EDFA then supplies a $\sim$30~dB gain to compensate for the loss to be incurred on the return channel to Alice. The total device insertion loss at Bob's terminal before the EDFA is 4.7\,dB. The amplified broadband light is phase modulated by an electro-optic modulator (EOM) using binary-phase shift keying with a 7\,Gbit/s pseudorandom sequence provided by a bit-error-rate tester. The EOM's embedded polarizer blocks the cross-polarized noise photons. Bob sends the amplified and modulated light back to Alice through another fiber channel.  

Placing the EOM after the EDFA overcomes the EOM's insertion loss. However, the amplified input to the EOM means that even a small amount of residual amplitude modulation can potentially allow Eve to obtain the bit values by simply measuring amplitude variations in the light Bob sends to Alice. A solution to overcoming the EOM's insertion loss without this vulnerability is to sandwich the EOM between two EDFAs, with the first EDFA having just enough gain to make the insertion loss inconsequential.   Residual amplitude modulation (typically no more than a few \%) will then be quite weak relative to the shot noise that will be seen by Eve.

Alice fine tunes the signal's round-trip propagation delay using a free-space delay line so that it matches the length of the fiber spool that stores her LO. In the homodyne measurement, the ASE portion of the returned light interferes with the retained LO at a 50:50 coupler, whose two outputs are fed to a balanced receiver (Discovery Semiconductors) with 14-GHz bandwidth and $\sim$55\% quantum efficiency. The balanced receiver's electrical output signal is amplified by a 44-dB-gain low-noise amplifier. The photodetection noise power primarily stems from the ASE photons created by Bob's EDFA; it is $>$15 dB higher than the shot-noise limit and $>$25 dB higher than the electrical noise floor, as confirmed by an electrical spectrum analyzer. Moreover, the electrical noise power is proportional to the signal power owing to Bob's EDFA. Hence, FL-QKD {\em does not} require a high-efficiency shot-noise limited homodyne measurement \cite{Zhuang2016}, whereas traditional CV-QKD does. 

A power splitter divides the electrical signal into two outputs. One is directed to the bit-error-rate tester, and the other output is rectified and fed to a lock-in amplifier for compensating the phase drift caused by differential thermal fluctuations between the signal and the reference paths. To do so, we implement a dither-and-lock servo using a 102\,kHz sinusoidal dither generated by the lock-in amplifier. The servo's error signal at the lock-in amplifier output is first processed by a PID-controller, then summed with the dither, before being combined with the EOM's driving input at a bias-T and fed to the EOM. 

Two essential parameters are needed to determine the SKR: Alice's bit-error probability $P_e$ and Eve's injection fraction $f_E$. We measured $P_e$ at different transmitted photons per bit with fixed source bandwidth. In each case, a set of 10 consecutive $P_e$'s were recorded and their standard deviations calculated. We measured $f_E$ using the channel monitor comprised of the three SNSPDs and a time-tagging device (Picoquant Hydraharp). The idler photons' detection times served as the synchronization signal for the Hydraharp's histogram measurements. The detection times of Alice's and Bob's tapped photons were recorded by two Hydraharp channels. Choosing a coincidence window of 3.2\,ns, the histograms allow us to determine the singles rate at Alice's tap $S_A$, the singles rate at Bob's tap $S_B$, the time-aligned (time-misaligned) coincidence rate between Alice's tap and the SPDC's idler $C_{IA}$ ($\widetilde{C}_{IA}$), and the time-aligned (time-misaligned) coincidence rate between Bob's tap and the SPDC's idler $C_{IB}$ ($\widetilde{C}_{IB}$). The injection fraction $f_E$ is then determined from \cite{Zhuang2016,supp} 
\begin{equation}
\label{eq:f_E}
	f_E = 1-\frac{C_{IB}-\widetilde{C}_{IB}}{C_{IA}-\widetilde{C}_{IA}}\times\frac{S_A}{S_B}.
\end{equation}
As discussed in \cite{Zhuang2016} and verified in \cite{Zhang2017}, $f_E$ is independent of the SNSPDs' quantum efficiencies, the channel transmissivity, and the source brightness. The channel monitor is thus calibration free. We performed 54 30-minute histogram measurements, giving $f_E = (0.053 \pm 0.066)\%$. After accounting for a 3-$\sigma$ confidence level for the measurement uncertainty, we obtained the upper bound $f_E^{\rm UB} = 0.3\%$ on Eve's injection fraction.

\begin{figure}[h]
\centering
\includegraphics[width=\linewidth]{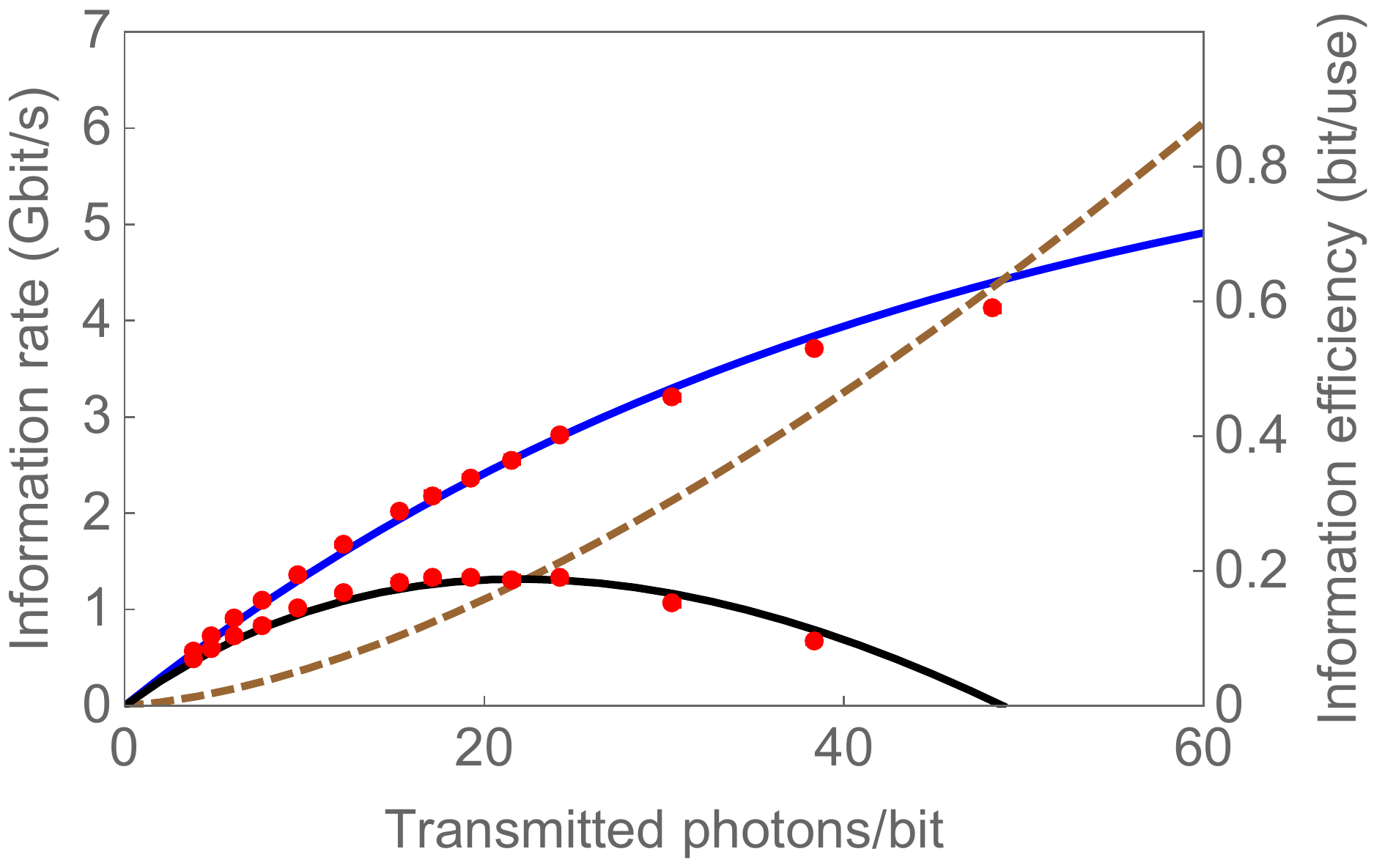}
\caption{\label{fig:SKRs} SKR (solid black) in Gbit/s, and $I_{AB}(P_e)$ (solid blue) and $\chi_{BE}(f_E)$ (dashed brown) in bits/use versus Alice's transmitted photons per bit.  Solid red circles are experimental values, with error bars ($\pm$1 standard deviation) that are mostly covered by the red circles.}
\label{fig:rate_3sigma}
\end{figure}

With $P_e$'s and $f_E$ measured, the SKRs are derived from
\begin{equation}
	{\rm SKR} = R\left[\beta I_{AB}(P_e) -\chi_{BE}(f_E^{\rm UB})\right].
\end{equation}
Here, $R = 7$\,Gbit/s is the modulation rate, $\beta = 0.94$ is the assumed reconciliation efficiency, $I_{AB}(P_e) = 1+P_e\log_2(P_e)+(1-P_e)\log_2(1-P_e)$ is Alice and Bob's bits/use mutual information, and $\chi_{BE}(f_E^{\rm UB})$ is the upper bound on Eve's bits/use Holevo information set by $f_E^{\rm UB}$ \cite{Zhuang2016}. We plot in Fig.~\ref{fig:SKRs} the theoretical curves for $I_{AB}(P_e)$ (solid blue), $\chi_{BE}(f_E^{\rm UB})$ (dashed brown), and the SKRs (solid black) at different transmitted photons per bit. The experimentally measured values are shown as red circles. The experimental measurements agree well with theory at transmitted photons per bit $<$ 30, where the phase-locking servo loop was optimized. At higher levels, the experimental data are slightly lower than the theoretical model due to sub-optimum servo loop operation. The maximum SKR is $1.3$\,Gbit/s, obtained at $\sim$20 transmitted photons per bit---or, equivalently, $\sim$$5.8\times10^{-4}$\,bit per optical mode obtained at $\sim$0.06\,photon per optical mode---over a 10-dB-attenuation channel for which $-\log_2(1-\eta) = 0.15$\,bit/mode.

In conclusion, we have demonstrated FL-QKD with 1.3\,Gbit/s SKR. This high SKR is made possible by FL-QKD's multi-mode encoding scheme in a two-way communication system, which enables transmission of many photons per bit to overcome channel loss while maintaining security. Future investigations along this line would encompass demonstrating FL-QKD in a field test, studying finite-size effects, and developing broadband QKD protocols in a networked setting.

This work was funded by the Air Force Office of Scientific Research (AFOSR) under Grant Number FA9550-14-1-0052 and the Office of Naval Research (ONR) under Contract Number N00014-16-C-2069. The authors thank S. W. Nam, D. Zhu, and Q. Zhao for support on SNSPDs, and E.~Wong for generating the three-dimensional experimental schematic.\\

\begin{widetext}

\begin{center}
\noindent{\Large\bf Supplemental Material}\\
\end{center}

We monitored the Alice-to-Bob channel to measure Eve's injection fraction, $f_E$, in order to bound Eve's Holevo information. The measured $f_E$ is calculated from \cite{Zhuang2016}
\begin{equation}
f_E = 1- \frac{C_{IB}-\tilde{C}_{IB}}{C_{IA}-\tilde{C}_{IA}}\times\frac{S_A}{S_B},
\label{feeqn}
\end{equation}
where $S_A$ is the singles rate at Alice's tap, $S_B$ is the singles rate at Bob's tap, $C_{IA}$ ($\widetilde{C}_{IA})$ is the time-aligned (time-misaligned) coincidence rate between Alice's tap and the SPDC's idler, and $C_{IB}$ ($\widetilde{C}_{IB}$) is  the time-aligned (time-misaligned) coincidence rate between Bob's tap and the SPDC's idler. Time-aligned coincidences are due to both the SPDC's signal photons and Alice's ASE light, whereas the time-misaligned coincidences result only from ASE light.

We performed channel-monitoring measurements under the operating conditions that gave the 1.3\,Gbit/s SKR:  SPDC signal and idler powers of $\sim$1.28\,pW after Alice's CWDM, a received ASE power of 1.47\,nW before Bob's EDFA, and no light injection by Eve. Our WSi SNSPDs begin to saturate at $\sim$1\,Mcounts/s, and have  $\sim$4\,Mcounts/s maximum count rates. Their effective quantum efficiencies drop, due to blocking loss, as they are pushed deeper into their saturation regimes. Our channel-monitor measurement duration was therefore primarily determined by the SNSPDs' maximum count rate. For the SPDC's idler counts, the effective quantum efficiency was $\sim$40\% at $\sim$4\,Mcounts/s.   Alice and Bob's tapped light is ASE plus a much weaker signal component from Alice's SPDC, so we attenuated the tapped beams prior to detection to prevent deeply saturating the SNSPDs.  Using $\sim$29\,dB (20\,dB) attenuation for Alice's (Bob's) tap, we kept the singles rates comparable to that for Alice's idler channel. Note that timing jitter increases in the SNSPDs' saturation regime. Hence, to count all SPDC coincidences, it is necessary to choose an appropriate coincidence window.

\begin{figure}[htb]
\centering
\includegraphics[scale=0.38]{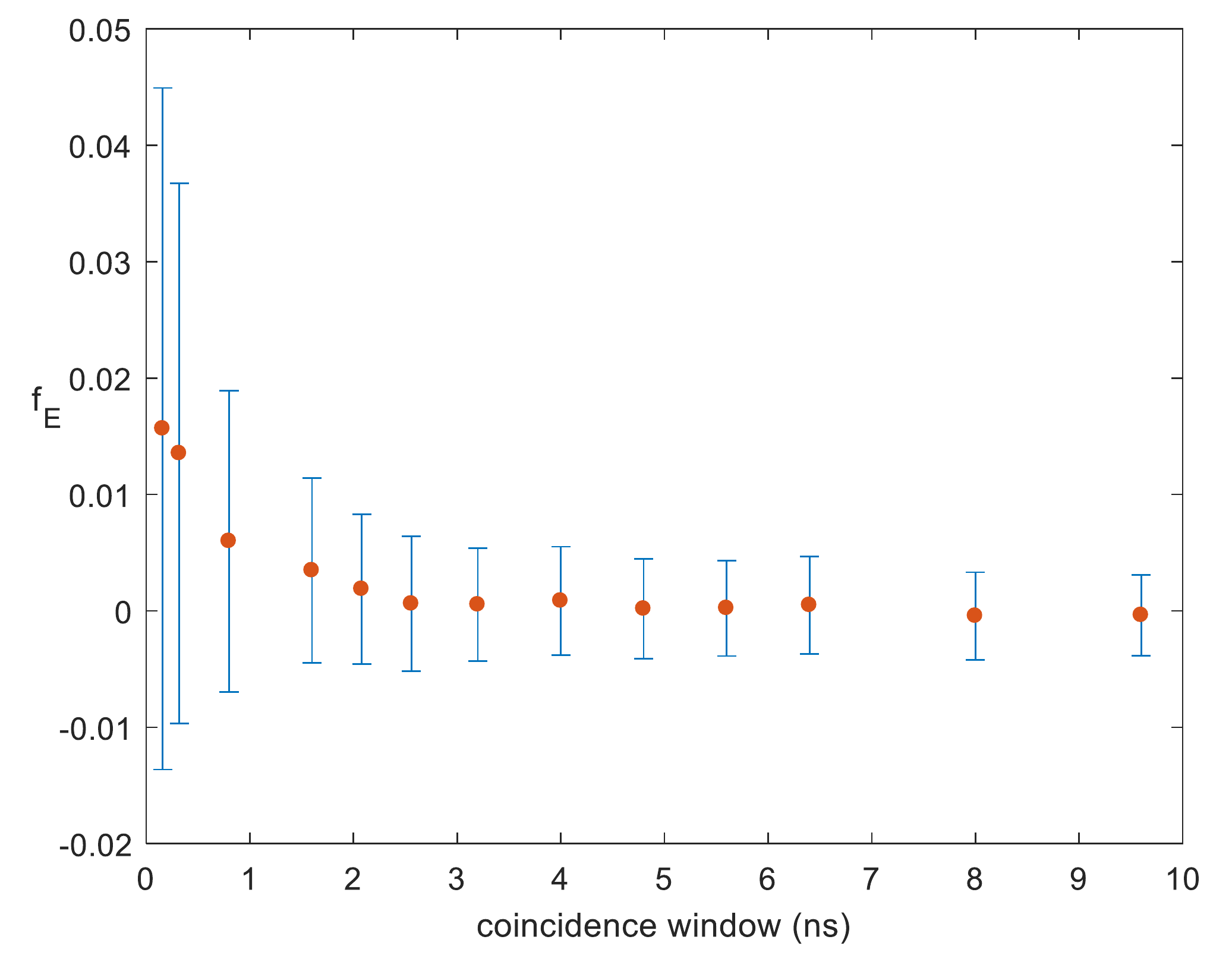}
\caption{Calculated values for Eve's injection fraction, $f_E$, versus coincidence-window duration. Each solid red circle is the average of 54 30-minute measurements, and the associated error bars are the estimated $\pm$\,1 standard deviation for a 30-minute measurement.}
\label{cwindowgraph}
\end{figure}
Operating our channel monitor's SNSPDs at $\sim$4\,Mcounts/s meant that we needed 54 30-minute-long measurements to obtain adequate signal-to-noise ratio.  Our $f_E$ values---calculated from the channel monitor's detection-time data using Eq.~(\ref{feeqn})---are plotted in Fig.~\ref{cwindowgraph} versus the coincidence-window duration.   
Decreasing the coincidence window below 3.2\,ns leads to uncounted SPDC coincidences, resulting in increasing $f_E$ estimates, despite there being no light injection by Eve. Coincidence windows between 3.2\,ns and 10\,ns, however, yielded $f_E$ estimates close to the true value with similar standard deviations. Therefore, to faithfully estimate $f_E$, we chose a 3.2\,ns coincidence window and, combining all 54 30-minute measurements, we obtained $f_E = (0.053\pm0.066)\%$ for the estimated injection fraction, where the uncertainty is $\pm 1$ standard error, i.e.,  the standard deviation for a 30-minute measurement divided by $\sqrt{54}$. 

The measurement time needed to obtain the preceding channel-monitor performance was quite long, but it can be significantly reduced if high-count-rate SNSPDs are employed. For example, the SNSPDs used in~\cite{shaw2014receiver} have 14\,Mcounts/s maximum count rates and 20\% effective quantum efficiencies. Using these detectors, we estimate that our 30-minute-long measurement time can be reduced to 25\,s, by pumping our SPDC crystal $\sim$7$\times$ harder to realize a 14\,Mcounts/s idler-beam singles rate.

\end{widetext}

\end{document}